\documentclass[12pt]{article}

\usepackage{amssymb}
\usepackage{amsmath}

\begin{document}

\title{\bf Eikonal Amplitude in the Gravireggeon Model at Superplanckian Energies}


\author{A.V. Kisselev\thanks{E-mail: alexandre.kisselev@mail.ihep.ru} \
and V.A. Petrov\thanks{E-mail: vladimir.petrov@mail.ihep.ru} \\
\small Institute for High Energy Physics, 142281 Protvino, Russia}

\date{}

\maketitle

\thispagestyle{empty}

\begin{abstract}
The gravity effects in high-energy scattering, described by a
four-dimensional eikonal amplitude related to gravireggeons
induced by compact extra dimensions are studied. It is
demonstrated that the real part of the eikonal (with a massless
mode subtracted) dominates its imaginary part at both small and
large impact parameters, in contrast to the usual case of hadronic
high-energy behavior. The real part of the scattering amplitude
exhibits an exponential falloff at large momentum transfer,
similar to that of the imaginary part of the amplitude.
\end{abstract}
\vfill \eject

\section{Introduction}
\label{sec:intr}

In our previous paper~\cite{Kisselev:03}, we considered the model
with compact extra spatial dimensions~\cite{Arkani-Hamed:98} and
calculated the contribution of Kaluza-Klein (KK) gravireggeons
into the inelastic cross section of high energy scattering of
four-dimensional SM particles. In particular, an expression for
the imaginary part of the eikonal has been derived. The results
were applied to cosmic neutrino gravitational interaction with
atmospheric nucleons~\cite{Kisselev:03}.

In the present paper, we study quantum gravity effects related to
the extra dimensions in the \emph{real part of the eikonal}. As in
Ref.~\cite{Kisselev:03}, the SM particles are confined on a
four-dimensional brane, while the gravity lives in all $D = d + 4$
dimensions. The extra dimensions are compactified with a radius
$R_c$. Thus, a fundamental mass scale, $M_D$, is related to the
Planck scale by a relation $ M_{Pl}^2  = M_D^{d+2}\, (2\pi
R_c)^d$~\cite{Arkani-Hamed:98}.

In the next Section we consider a case of one extra dimensions.
The generalization to more than one extra dimension is given in
Section~3. The conclusions and discussions are given in the last
Section. Some technical details of our calculations are collected
in Appendices.

\section{One extra dimension ($\mathbf{d = 1}$)}
\label{sec:d=1}

For the sake of simplicity and for pedagogical reasons, we will
consider first one extra dimension. The general case ($d>2$) will
be analyzed in the next section. In the gravireggeon model,
eikonal is given by the sum of reggeized KK gravitons in the
$t$-channel~\cite{Kisselev:03}:
\begin{equation}
\chi(s,b) = \frac{1}{8\pi s} \int\limits_{-\infty}^{0} dt \, J_o(b
\sqrt{-t}) \, \sum_{n= - \infty}^{\infty}  A^B(s, t, n),
\label{10}
\end{equation}
where $\sqrt{s}$ is an invariant energy, and the Born amplitude is
of the form
\begin{equation}
A^B(s,t,n) = G_N \Big[ i - \cot \frac{\pi}{2} \alpha_n(t) \Big]
\alpha_g' \, \beta^2_n(t) \left( \frac{s}{s_0} \right)^{
\alpha_n(t)}.
\label{12}
\end{equation}
Here $n$ is a KK-number. The value $n=0$ corresponds to usual
massless graviton.

Both massless graviton and its KK massive excitations lie on
linear Regge trajectories:
\begin{equation}
\alpha(t_D) = \alpha(0) + \alpha'_g \, t_D,
\label{14}
\end{equation}
where $t_D$ denotes $D$-dimensional momentum transfer. Since the
extra dimension is compact with the radius $R_c$, we come to
splitting of the Regge trajectory \eqref{14} into a leading vacuum
trajectory,
\begin{equation}
\alpha_0(t) \equiv \alpha_{grav}(t) = 2 + \alpha_g' t,
\label{16}
\end{equation}
and infinite sequence of secondary, ``KK-charged'',
gravireggeons~\cite{Petrov:02}:
\begin{equation}
\alpha_n(t) = 2 - \frac{\alpha_g'}{R_c^2}\,n^2 + \alpha_g' t,
\quad n \geqslant 1.
\label{18}
\end{equation}
The string theory implies that the slope of the gravireggeon
trajectory is universal for all $s$, and $\alpha_g' = 1/M_s^2$,
where $M_s$ is a string scale.

In ref.~\cite{Kisselev:03} the imaginary part of the
eikonal~\eqref{10} has been calculated. In the present paper we
consider the real part of the eikonal. From Eqs.~\eqref{10},
\eqref{12} and \eqref{18} we obtain ($q_{\bot}^2 = -t$):
\begin{eqnarray}
\text{Re}\,\chi(s,b) &=& G_N s \frac{\alpha_g'}{8}
\int\limits_0^{\infty} q_{\bot} d q_{\bot} \, J_0(q_{\bot} b) \,
e^{- q_{\bot}^2 R_g^2(s)}
\nonumber \\
&\times& \sum_{n= - \infty}^{\infty} \, \cot \big[ \frac{\pi
\alpha_g'}{2} \big( -t + \frac{n^2}{R_c^2} \big) \big] e^{- n^2
R_g^2(s)/ R_c^2},
\label{20}
\end{eqnarray}
where
\begin{equation}
R_g(s) = \sqrt{\alpha_g' \ln (s/s_0)}
\label{22}
\end{equation}
is a gravitational slope (dynamical radius). Formally, there exist
poles in the sum in Eq.~\eqref{20} at negative values of
$\alpha_n(t)$. It is demonstrated in Appendix~A that these tachyon
poles are fictitious singularities, and, thus, they will not be
taken into account in our calculations.

In what follows, we will assume that $t$ lies in the physical
region, $t<0$, and
\begin{equation}
\alpha_g' \, |t| \ll 1.
\label{24}
\end{equation}
It is equivalent to $|t| \ll M_s^2$, where the string scale $M_s$
is of order 1 TeV. The sum in \eqref{20} is effectively cut from
above, $n \lesssim n_{\max} = R_c /R_g(s)$. It means that
$\alpha_g' n^2/R_c^2 \lesssim [\ln (s/s_0)]^{-1} \ll 1$ in
\eqref{20}.

Let us  define $\text{Re}\,\check{\chi} (s,b)$ to be the real part
of the eikonal with a pole term (corresponding to $n=0$ in
\eqref{20}) subtracted. With all mentioned above, it can be
written as follows:%
\footnote{Taking into account that effectively $|\alpha_n(t) - 2|
\ll 1$.}
\begin{eqnarray}
\text{Re}\,\tilde{\chi} (s,b) &=& G_N R_c^2 \,  s \frac{1}{2 \pi}
\int\limits_0^{\infty} q_{\bot} d q_{\bot} \, J_0(q_{\bot} b) \,
e^{- q_{\bot}^2 R_g^2(s)}
\nonumber \\
&\times& \sum_{n=1}^{\infty} \, \frac{1}{n^2 + R_c^2 |t|} \, e^{-
n^2 R_g^2(s)/ R_c^2}.
\label{26}
\end{eqnarray}

One can see that $\varepsilon (s) = R_g(s)/R_c \ll 1$ even at
ultra-high (cosmic) energies $s$. Indeed, a magnitude of
$\varepsilon (s)$ is defined by the ratio $\sim M_c/M_s$, with a
compactification mass scale, $M_c = R_c^{-1}$, varying from
$10^{-3}$ eV for $d=2$ to $10$ MeV for $d = 6$. So, $\varepsilon
(s)$ is taken to be a small parameter everywhere in our
calculations.

Let us consider two distinct regions of the momentum transfer
$|t|$. For $0 \leqslant |t| \ll  R_c^{-2}$, the leading term looks
like
\begin{equation}
I_1 = \sum_{n=1}^{\infty} \, \frac{1}{n^2 + R_c^2 |t|} \, e^{- n^2
R_g^2(s)/ R_c^2} \Big|_{|t|R_c^2 \ll 1} \simeq \frac{\pi^2}{6} -
\frac{\pi^4}{90} \, R_c^2 \, |t| + \mathrm{O} (\varepsilon (s)).
\label{28}
\end{equation}
At large $|t|$ ($R_c^{-2} \ll |t| < \infty$), we will consider two
subregions. If the momentum transfer runs the subregion $R_c^{-2}
\ll |t| \ll R_g^{-2}(s)$, then
\begin{equation}
I_1 = \sum_{n=1}^{\infty} \, \frac{1}{n^2 + R_c^2 |t|} \, e^{- n^2
R_g^2(s)/ R_c^2} \Bigg|_{\genfrac{}{}{0pt}{}{|t|R_c^2 \gg
1}{|t|R_g^2(s) \ll 1}} \simeq \frac{\pi}{2R_c \sqrt{|t|}} -
\frac{1}{R_c^2|t|} + \mathrm{O} (\varepsilon (s)).
\label{30}
\end{equation}
Note, the leading terms in \eqref{28} and \eqref{30} match at
$R_c\sqrt{|t|} = a_1 = 3/\pi$. At very large values of $|t|$,
namely, for $R_g^{-2}(s) \lesssim |t| < \infty$, the sum in
\eqref{30} has the asymptotics
\begin{equation}
I_1 \Big|_{|t|R_g^2(s) \gg 1} \simeq \frac{\pi}{2R_c R_g(s) |t|}.
\label{31}
\end{equation}

It can be shown that a contribution from the region $R_g^{-2}(s)
\lesssim |t| < \infty$ is suppressed as compared to the region
$R_c^{-2} \lesssim |t| < \infty$ by the factor $ \sim \varepsilon
(s)$ (for large impact parameter which we are interested in).
Thus, we can write (using table integrals from
~\cite{Prudnikov:II}):
\begin{eqnarray}
&& \text{Re}\,\tilde{\chi} (s,b) \simeq G_N R_c^2 s \, \Big[
\frac{\pi}{12} \, \int\limits_0^{\ a_1 R_c^{-1}} q_{\bot} d
q_{\bot} \, J_0(q_{\bot} b) \, e^{- q_{\bot}^2 R_g^2(s)}
\nonumber \\
&& + \frac{1}{4R_c} \, \int\limits_{a_1 R_c^{-1}}^{\infty} d
q_{\bot} \, J_0(q_{\bot} b) \, e^{- q_{\bot}^2 R_g^2(s)} \Big]
\simeq \frac{G_N R_c s}{4} \, \left\{ \frac{1}{b} \, J_1 \left(
\frac{a_1 b}{R_c} \right) + \frac{a_1}{R_c} \right.
\nonumber \\
&& \times \left. \left[ \frac{\sqrt{\pi} R_c}{2 a_1 R_g(s)} \,
\Phi \left( \frac{1}{2}, \, 1; \, - \frac{b^2}{4 R_g^2(s)} \right)
- \, {}_1F_2 \left( \frac{1}{2}; \, 1, \, \frac{3}{2}; \, -
\frac{a_1^2 b^2}{4R_c^2} \right) \right] \right\},
\label{32}
\end{eqnarray}
where $\Phi(a;b;z)$ is the confluent hypergeometric function,%
\footnote{The confluent hypergeometric function $\Phi
(1/2,\,1;\,z)$ can be related to the modified Bessel function
$I_0(z/2)$.}
and ${}_1F_2(a;b,c;z)$ is the generalized hypergeometric
function~\cite{Erdelyi:I}. For $b \gg R_c^2/R_g(s) \gg R_c$, we
get the following asymptotics~\cite{Erdelyi:I}:
\begin{equation}
\left. \frac{\sqrt{\pi} R_c}{2a_1 R_g(s)} \, \Phi \left(
\frac{1}{2}, \, 1; \, - \frac{b^2}{4R_g^2(s)} \right) \right|_{b
\varepsilon(s)\gg R_c} \!\! \simeq \frac{R_c}{a_1 b} \Big[ 1 +
\mathrm{O} \left( \frac{R_g^2(s)}{b^2} \right) \Big],
\label{36}
\end{equation}
and%
\footnote{The generalized hypergeometric function  ${}_1F_2
(1/2;\,1,\,3/2;\, - z)$ can be defined by the integral of the
Bessel function $J_0(z)$. As a result, its asymptotics has
oscillations.}
\begin{equation}
\left. {}_1F_2 \left( \frac{1}{2}; \, 1, \, \frac{3}{2}; \, -
\frac{a_1^2 b^2}{4R_c^2} \right) \right|_{b \gg R_c} \simeq
\frac{R_c}{a_1 b} \Big[ 1 + J_1 \left( \frac{a_1 b}{R_c} \right) +
\frac{R_c}{a_1 b} \, J_2 \left( \frac{a_1 b}{R_c} \right) \Big].
\label{38}
\end{equation}
Here $J_1(z)$ and  $J_2(z)$ are the Bessel functions. As a result,
we obtain from Eqs.~\eqref{32} and \eqref{36}-\eqref{38}:
\begin{equation}
\text{Re} \tilde{\chi}(s,b) \Big|_{b \gg R_c} \simeq - G_N s
\frac{\pi}{12} \left( \frac{R_c}{b} \right)^{2} \! J_2 \left(
\frac{a_1 b}{R_c} \right), \label{40}
\end{equation}
where the constant $a_1$ is defined after formula \eqref{30}.

At zero impact parameters, one has
\begin{equation}
\text{Re}\,\tilde{\chi} (s,b=0) \sim G_N s \, \frac{R_c}{R_g(s)}.
\label{41}
\end{equation}

On the other hand, the imaginary part of the eikonal was found to
be (for $d=1$) \cite{Kisselev:03}
\begin{equation}
\text{Im}\,\chi(s,b) = \sqrt{\pi} \, G_N s \frac{R_c}{R_g(s)}
\big[ \ln (s/s_0) \big]^{-1} \exp [ -b^2 /4 R_g^2(s)].
\label{42}
\end{equation}
So, the real part of the eikonal dominates the imaginary part at
zero impact parameter,%
\footnote{We remind that the singular term was subtracted in
$\text{Re}\,\tilde{\chi}$.}
$\text{Re} \, \tilde{\chi} (s,0) / \text{Im} \, \tilde{\chi} (s,0)
\sim \ln s$, and it has a power-like behavior (with oscillations)
at $b \rightarrow \infty $ \eqref{40}, while the imaginary part
decreases exponentially at large $b$.

\section{More than two extra dimensions ($\mathbf{d > 2}$)}
\label{sec:d>2}

The expression for the real part of the eikonal \eqref{26} is
easily generalized for $d>2$:
\begin{eqnarray}
\text{Re}\,\tilde{\chi} (s,b) &=& G_N R_c^2 s \, \frac{1}{2 \pi}
\int\limits_0^{\infty} q_{\bot} d q_{\bot} \, J_0(q_{\bot} b) \,
e^{- q_{\bot}^2 R_g^2(s)}
\nonumber \\
&\times& \sum_{n=1}^{\infty} \, \frac{1}{n^2 + R_c^2 |t|} \, e^{-
n^2 \varepsilon^2(s)} \!\!\! \!\!\!\sum_{n_1^2 + n_2^2 + \cdots
n_{d-1}^2 \leqslant \, n^2}\!\!\!\!\!,
\label{100}
\end{eqnarray}
where the notation $n^2=\sum_i^d n_i^2$ is introduced. The main
contribution to the sum in the RHS of Eq.~\eqref{100} comes from
the region $n^2 \sim (d-2)/\varepsilon^2(s) \gg 1$. Thus, to
estimate the sum in $(n_1, n_2, \ldots n_{d-1})$ analytically, we
can replace the sum by the integral:
\begin{eqnarray}
I_d &=& \sum_{n=1}^{\infty} \, \frac{1}{n^2 + R_c^2 |t|} \, e^{-
n^2 \varepsilon^2(s)} \!\!\! \!\!\!\sum_{n_1^2 + n_2^2 + \cdots
n_{d-1}^2 \leqslant \, n^2}
\nonumber \\
&\rightarrow& \, \sum_{n=1}^{\infty} \, \frac{1}{n^2 + R_c^2 |t|}
\, e^{- n^2 \varepsilon^2(s)} \idotsint\limits_{\vec{x}^2
\leqslant n^2} \!\!  d\vec{x}
\nonumber \\
&=& \frac{\pi^{(d-1)/2}}{\Gamma \big( \frac{d+1}{2} \big)} \,
\sum_{n=1}^{\infty} \, \frac{ n^{d-1}}{n^2 + R_c^2 |t|} \,\, e^{-
n^2 \varepsilon^2(s)}.
\label{120}
\end{eqnarray}

As in the previous section, we consider two regions of the
momentum transfer. At $|t| \ll  R_c^{-2}$ we obtain (up to
corrections $\mathrm{O}(R_c^2|t|)$ and
$\mathrm{O}(\varepsilon^2(s))$):
\begin{eqnarray}
I_d &\simeq& \sum_{n=1}^{\infty} \, \frac{1}{n^2} +
\frac{\pi^{(d-1)/2}}{\Gamma \big( \frac{d+1}{2} \big)} \,
\sum_{n=1}^{\infty} \, \frac{1}{n^2 + R_c^2 |t|} \, (n^{d-1} - 1)
\,\, e^{-n^2 \varepsilon^2(s)}
\nonumber \\
&\rightarrow& \frac{\pi}{6} +  \frac{\pi^{(d-1)/2}}{2\Gamma \big(
\frac{d+1}{2} \big)} \, \int\limits_{1}^{\infty} dz \, (z^{d/2-2}
- z^{-3/2}) \, e^{- z \varepsilon^2(s)}
\nonumber \\
&\simeq& \frac{\pi}{6} +  \frac{\pi^{(d-1)/2}}{2\Gamma \big(
\frac{d+1}{2} \big)} \, \left[ \Psi \Big(
1,\,\frac{d}{2};\,\varepsilon^2(s) \Big) - \Psi \Big(
1,\,\frac{1}{2};\,\varepsilon^2(s) \Big)\right],
\label{140}
\end{eqnarray}
where $\Psi(a,b;z)$ is the confluent hypergeometric function,%
\footnote{The functions  $\Psi(a,b;z)$ and  the above  mentioned
function $\Phi(a,b;z)$ are different solutions of the
confluent hypergeometric equation.}
and we have replaced the sum in $n$ by the integral.%
\footnote{Note, for $d=1$ this is justified only if $|t| R_c^2 \gg
1$.}
For $d>2$, Eq.~\eqref{140} results in
\begin{equation}
I_d \Big|_{|t|R_c^2 \ll 1} \simeq \frac{\pi^{(d-1)/2} \Gamma \big(
\frac{d}{2} - 1 \big)}{2\Gamma \big( \frac{d+1}{2} \big)} \left(
\frac{1}{\varepsilon(s)} \right)^{d-2},
\label{160}
\end{equation}
neglecting insignificant terms $\mathrm{O}(\varepsilon (s))$.
Starting from Eq.~\eqref{140}, we come to the asymptotics $I_1
\simeq \pi^2/6$, in accordance with Eq.~\eqref{28}.

Now let us consider the region $R_c^{-2} \ll |t| < \infty $. Then
the quantity $I_d$ can be cast in the form:
\begin{eqnarray}
I_d &\simeq& \frac{\pi^{(d-1)/2}}{2\Gamma \big( \frac{d+1}{2}
\big)} \, \int\limits_{1}^{\infty} dz \, \frac{z^{d/2-1}}{z +
R_c^2 |t|} \,\, e^{- z \varepsilon^2(s)}
\nonumber \\
&\simeq& \frac{\pi^{(d-1)/2}}{2\Gamma \big( \frac{d+1}{2} \big)}
\, \left[ \Gamma \big( \frac{d}{2}\big) (R_c^2|t|)^{d/2-1} \, \Psi
\Big( \frac{d}{2},\,\frac{d}{2};\, R_g^2(s)|t| \Big) - \frac{2}{d}
\frac{1}{R_c^2|t|} \right].
\label{180}
\end{eqnarray}
For $d>2$, we should consider separately two subregions. Namely,
if the momentum transfer is bounded by inequalities $R_c^{-2} \ll
|t| \ll R_g^{-2}(s)$, Eq.~\eqref{180} results in previously
obtained asymptotics \eqref{160}. At very large values of $|t|$,
such as $R_g^{-2}(s) \ll |t| < \infty$, one gets from \eqref{180}:
\begin{equation}
I_d \Big|_{|t|R_c^2 \gg 1} \simeq \frac{\pi^{(d-1)/2} \Gamma \big(
\frac{d}{2} \big)}{2\Gamma \big( \frac{d+1}{2} \big)} \, \left(
\frac{1}{\varepsilon(s)} \right)^{d-2} \! \frac{1}{R_g^2(s)|t|}.
\label{200}
\end{equation}
For $d=1$, the correct values of $I_1$ \eqref{30}, \eqref{31} are
reproduced. The asymptotics \eqref{180} and \eqref{200} match at
$|t| = a^2 R_g^{-1}(s)$, where
\begin{equation}
a^2 = (d-2)/2.
\label{210}
\end{equation}

Thus, we get the following expression for the eikonal:
\begin{eqnarray}
&& \text{Re}\,\tilde{\chi} (s,b) \simeq G_N s \, \left[
\frac{R_c}{R_g(s)} \right]^d \, \frac{\pi^{(d-3)/2} \Gamma \big(
\frac{d}{2} \big)}{4\Gamma \big( \frac{d+1}{2} \big)}
\nonumber \\
&& \times \Big[ \frac{R_g^2(s)}{a^2} \!\int\limits_0^{\ a
R_g^{-1}(s)} \!\!\! q_{\bot} d q_{\bot} \, J_0(q_{\bot} b) \, e^{-
q_{\bot}^2 R_g^2(s)} +  \, \int\limits_{a R_g^{-1}(s)}^{\infty}
\!\! \frac{d q_{\bot}}{ q_{\bot}} \, J_0(q_{\bot} b) \, e^{-
q_{\bot}^2 R_g^2(s)} \Big]
\nonumber \\
&& \equiv G_N  s \, \left[ \frac{R_c}{R_g(s)} \right]^d \,
\frac{\pi^{(d-3)/2} \Gamma \big( \frac{d}{2} \big)}{4\Gamma \big(
\frac{d+1}{2} \big)} \, \Big[  I_< \left( \frac{a b}{ R_g(s)}
\right) + I_> \left( \frac{a b}{ R_g(s)} \right) \Big].
\label{220}
\end{eqnarray}

At small impact parameter, we get immediately from \eqref{220}:
\begin{eqnarray}
\text{Re}\,\tilde{\chi} (s,b)\Big|_{b \ll R_g(s)} &=& C(d) \, G_D
s \, \alpha_g'^{-d/2} \, \Big[ \ln \Big( \frac{s}{s_0} \Big)
\Big]^{-d/2}
\nonumber \\
&+ &  \mathrm{O} \left( \frac{b^2}{R_g^2(s)} \right),
\label{240}
\end{eqnarray}
where $C(d)$ is a constant depending on the number of the extra
dimensions, explicit form of which can be obtained from
\eqref{220}. The asymptotics of the real part of the eikonal at
large $b$ is more complicated to analyze. For $b \gg R_g(s)$, it
is calculated in the Appendix~B, and the leading term looks like
\begin{eqnarray}
\text{Re}\,\tilde{\chi} (s,b) \Big|_{b \gg R_g(s)} \!\! &\simeq& -
G_D s \, \alpha_g'^{-d/2} \, \, \frac{e^{-a^2} \, \Gamma \big(
\frac{d}{2} \big)}{\pi^{(d+3)/2} 2^{d+1} a^2 \Gamma \big(
\frac{d+1}{2} \big)} \, \, \Big[ \ln \Big( \frac{s}{s_0} \Big)
\Big]^{-d/2}
\nonumber \\\
&\times& \left( \frac{R_g(s)}{b} \right)^{2} \! J_2 \! \left(
\frac{a b}{R_g(s)} \right).
\label{260}
\end{eqnarray}

The expression for the imaginary part of the eikonal for $d
\geqslant 1$ was calculated in Ref.~\cite{Kisselev:03}:
\begin{eqnarray}
\text{Im} \, \chi(s,b) &=& \frac{ G_D s \,
\alpha_g'^{-d/2}}{\pi^{d/2 -1}} \, \Big[\ln \Big( \frac{s}{s_0}
\Big) \Big]^{-(1+d/2)}
\nonumber \\
&\times& \exp [ -b^2 /4 R_g^2(s)].
\label{280}
\end{eqnarray}
As one can see from \eqref{240} and \eqref{280},
\begin{equation}
\frac{\text{Re}\,\tilde{\chi} (s,0)}{\text{Im} \, \chi(s,0)} \sim
\ln s .
\label{300}
\end{equation}

Let us stress, we study the case when \emph{colliding particles
are confined on the 4-dimensional brane}, with gravity living in
all $D$ dimensions. We see that the real part of the eikonal in
$b$-space has a power-like behavior in $b$ with oscillations,
while the imaginary part decreases exponentially at $b \gg
2\alpha_g' \ln s$. Both depend on the Regge slope $\alpha_g'$ via
the gravitational radius $R_g(s)$ \eqref{22}.%
\footnote{It does not take place, if colliding particles live in
$D$ dimensions~\cite{Muzinich:88}.}

Let us remind the asymptotic behavior of the eikonal function
derived in the framework of the string theory for the scattering
of \emph{$D$-dimensional fields in a flat
space-time}~\cite{Amati:87}:
\begin{equation}
\chi_{_D}(s,b) \Big|_{b^2 \gg \alpha' \ln s} \simeq \left(
\frac{b_c}{b} \right)^d + i \pi^2 \frac{G_N^{^D} s \,
\alpha'^{-d/2}}{(\pi \ln s)^{1 + d/2}} \exp \! \Big( -
\frac{b^2}{4 \alpha' \ln s} \Big),
\label{320}
\end{equation}
where $b_c = [2 \pi^{-d/2} \Gamma(d/2) G_N^{^D} s ]^{1/d}$,
$G_N^{^D}$ being the Newton constant in $D$ flat dimensions. Note,
the real part of $\chi_{_D}(s,b)$ exhibits power-law falloff which
does not depend on the string tension $\alpha' \equiv \alpha_g'$.

One can observe, taking into account the definition of the
gravitational radius $R_g(s)$~\eqref{22}, that the \emph{imaginary
parts} of $\chi(s,b)$ and $\chi_{_D}(s,b)$ \emph{coincide} at $b
\gg \alpha_g' \ln s$. As for the real part of the eikonal,
$\text{Re}\,\tilde{\chi} (s,b)$ decreases as a \emph{fixed
(d-independent) power of} $b$ at $b \rightarrow \infty$, contrary
to \eqref{320}. The scales in the real parts (associated with the
impact parameter $b$) are also different: dynamical radius
$R_g(s)$ in our case, and $b_c \sim (G_N^{^D} s)^{1/d}$ in
$\chi_{_D}(s,b)$~\eqref{320}.

Because of the inequality $R_g(s) \ll R_c$, our formulae contain
the compactification radius $R_c$ only via $D$-dimensional
coupling $G_D = M_D^{-(2+d)} = G_N (2\pi R_c)^d$. However, at
extremely high energies, when the dynamical radius $R_g(s)$
becomes comparable with (or larger than) $R_c$, the eikonal
profile in impact parameter space should `` feel'' the size of the
compact dimensions $R_c$~\cite{Kisselev:03}.

In this connection, let us mention the SM in a
\emph{$D$-dimensional space-time} with compact extra dimensions,
but \emph{without gravity}~\cite{Petrov:01}. In such a case, the
dynamical radius, $R(s)$, is proportional to $\ln
(s/s_0)/\sqrt{t_0}$, where $t_0$ denotes the nearest (non-zero)
singularity in the $t$-channel (for instance, $t_0 = m_{\pi}^2$,
if only strong interactions are taken into account).

The expression for four-dimensional eikonal amplitude (in the
presence of $d$ compact extra dimensions) looks
like~\cite{Kisselev:03}:
\begin{equation}
A(s,t) = 2i s \int d^2 b \; e^{i q_{\bot} b} \left[ 1 - e^{i
\chi(s,b)}\right].
\label{340}
\end{equation}
At not extreme energies, namely, for $\sqrt{s} \lesssim M_D \sim
M_s$, we have inequalities $\text{Re} \, \tilde{\chi}(s,b), \,
\text{Im} \, \chi(s,b) \ll 1$, and Eq.~\eqref{340} is given by
\begin{eqnarray}
\tilde{A}(s,t) &\simeq& 4\pi s \int\limits_0^{\infty} db \, b \;
J_0(q_{\bot} b) \left[\text{Re} \, \tilde{\chi}(s,b) + i \,
\text{Im} \, \chi(s,b) \right]
\nonumber \\
&=& \text{Re} \tilde{A}(s,t) + i \, \text{Im} \, A(s,t).
\label{360}
\end{eqnarray}
The imaginary part of the scattering amplitude exhibits
exponential falloff at large $|t|$:
\begin{equation}
\text{Im} \, A(s,t) = \frac{8 \, G_D s^2 \,
\alpha_g'^{1-d/2}}{\pi^{d/2-2}} \, \Big[\ln \Big( \frac{s}{s_0}
\Big) \Big]^{-d/2} \exp \, \big(\, t \alpha_g' \ln (s/s_0) \,
\big).
\label{380}
\end{equation}

As for the real part of the amplitude, we obtain the following
behavior (see Appendix~C for details):
\begin{eqnarray}
&& \text{Re} \tilde{A}(s,t) =  G_D s^2 \, \alpha_g'^{-d/2} \,
\frac{\Gamma \big( \frac{d}{2} \big)}{2^d \pi^{(d+1)/2}\Gamma
\big( \frac{d+1}{2} \big)} \, \Big[ \ln \Big( \frac{s}{s_0} \Big)
\Big]^{-d/2}
\nonumber \\
&& \times  \left\{
\begin{array}{c}
  \displaystyle
  \frac{\alpha_g'
  \ln (s/s_0) }{a^2} \, \Big[ 2 + \frac{\alpha_g' \ln (s/s_0)  \, t}{a^2}
  \Big] + \frac{1}{t} \, \big[ 1 - \exp \, \big(\, t \alpha_g'
  \ln (s/s_0) \, \big) \big] \\
  \big (0 < \alpha'_g \, |t| < a^2/ \ln(s/s_0) \big) \\ \\
  \displaystyle
  \frac{1}{-t} \, \exp \, \big(\, t \alpha_g' \ln (s/s_0) \, \big)
  \\
  \big( \alpha'_g \,|t| \geqslant a^2/ \ln(s/s_0) \big) \\
\end{array}
\right.
\label{400}
\end{eqnarray}
Note that $\text{Im} \, A(s,t) \ll \text{Re} \, \tilde{A}(s,t)$ in
the kinematical region~\eqref{24}, in particular,
\begin{equation}
\frac{\text{Re} \, \tilde{A} (s,t)}{\text{Im} \, A(s,t)}
\Bigg|_{t=0} \sim \ln s , \qquad \frac{\text{Re} \, \tilde{A}
(s,t)}{\text{Im} \, A(s,t)} \Bigg|_{1 \gg \alpha'_g \,|t| \gg (\ln
s)^{-1}} \sim \frac{1}{\alpha_g' \, |t|}.
\label{410}
\end{equation}

The asymptotics of the amplitude at large $|t|$ \eqref{400} is
quite different from the behavior of the eikonal amplitude in both
the string theory~\cite{Amati:87} and in the model with Regge
exchanges in $D$ \emph{flat dimensions}~\cite{Muzinich:88}:
\begin{equation}
A(s,t) \Big|_{|t| \gg b_c^{-2}} \sim G_N^{^D} s^2 \,
\alpha_g'^{(1-d)/2} |t|^{-(d+2)^2/4(d+1)} \, e^{i \phi_D(t)},
\label{420}
\end{equation}
where $\phi_D(t) \sim |t|^{d/2(d+1)}$, and $b_c$ is define after
formula~\eqref{320}. Formula~\eqref{400} is also different from
the asymptotic behavior of $A(s,t)$ in the model with
\emph{compact extra dimensions}, when non-reggeized KK graviton
exchanges are summed up~\cite{Giudice:02}:
\begin{equation}
A(s,t)  \Big|_{|t| R_c^2 \gg 1} \sim G_D s^2 \,
\alpha_g'^{(1-d)/2} |t|^{-(d+2)/2(d+1)} \, e^{i \phi_D(t)}.
\label{440}
\end{equation}

It is worth to note that \eqref{440} decreases as a power of $|t|$
(the latter being larger than $-1$ for $d \geqslant 1$), in spite
of the fact that both amplitudes describe the scattering of fields
trapped \emph{on the brane}. This can be understood as follows.
For non-reggeized exchanges~\cite{Giudice:02}, the sum in KK
numbers $(n_1, \, n_2, \ldots n_d)$ contains no suppression factor
$\exp[- n^2 R_g^2(s)/ R_c^2]$, contrary to our approach with the
gravireggeon exchanges~\eqref{100}. The sum diverges and needs a
definition for $d>1$. Usually, the sum is replaced by a
$d$-dimensional integral, which is calculated by using dimensional
regularization. This procedure leads to the power-like falloff of
the eikonal with $d$-depending power, similar to the case when
colliding fields are not confined to the brane, but can propagate
in the extra dimensions~\cite{Amati:87}. Moreover, the eikonal is
pure real in this scheme~\cite{Giudice:02}.

\section{Conclusions}
\label{sec:concl}

In the framework of the model with $d$ extra compact dimensions,
we have calculated the quantum gravity effects related to the
gravireggeon exchanges in $t$-channel. For the scattering of the
SM fields living on the 4-dimensional brane, the real part of the
eikonal (with the massless mode subtracted) is estimated. It is
shown that it decreases as a power of $b$ (with oscillations) at
large values of the impact parameter $b$. This power does not
depend on the number of the extra dimensions $d$, contrary to the
case when the colliding fields are allowed to propagate in the
bulk. The scale, associated with the impact parameter $b$, is
$\alpha_g'^{-1}$, while in the $D$-dimensional flat space-time the
corresponding scale is defined by $(G_N^{^D} s)^{1/d}$, where
$G_N^{^D}$ is the Newton constant in $D$ dimensions.

The calculations complete our results on the imaginary part of the
eikonal obtained previously. In particular, it was shown that the
imaginary parts of the eikonal are the same for the case when
colliding particles are confined to the brane and when they
propagate freely in extra dimensions. In the present paper, we
have also calculated the eikonal amplitude and have shown, that
both the real part of the amplitude and its imaginary part
decreases exponentially at large momentum transfer.

The real part of the amplitude dominates the imaginary part at
zero momentum transfer, in contrast to high-energy behavior of
hadronic amplitudes (see, for instance, Ref.~\cite{Eden}). Note,
however, that this result was obtained in the region $\ln s \ll
R_c^2/\alpha'_g$. At asymptotical $s$, the inequality $|\text{Re}
\, A (s,0)|/|\text{Im} \, A(s,0)| < \text{const}$ will be
reproduced, provided the massless mode is discarded in the
eikonal.

\setcounter{equation}{0}
\renewcommand{\theequation}{A.\arabic{equation}}

\section*{Appendix A}
\label{sec:AppA}

The Sommerfeld-Watson transformation results in the following
expression for a contribution of the Regge trajectory $\alpha(t)$
to the amplitude~\cite{Collins}:
\begin{eqnarray}
&& A(s,t)\Big|_{pole} = -16 \pi^2 [2\alpha(t) + 1] \beta(t) \nonumber \\
&& \times \left[ \frac{1 + \xi \exp(- i \pi \alpha(t))}{\sin \pi
\alpha(t)} \, P_{\alpha(t)}(-z_t(s,t)) - \xi \frac{2}{\pi} \,
Q_{\alpha(t)}(-z_t(s,t)) \right].
\label{A02}
\end{eqnarray}
Here $\xi$ is a signature of the trajectory, $z_t(s,t)$ is a
cosine of a scattering angle in the $t$-channel, $\beta(t)$ is a
residue of the Regge pole in a partial amplitude:
\begin{equation}
A_l^{\xi}(t) \Big|_{l \rightarrow \alpha} \simeq \frac{\beta(t)}{l
- \alpha(t)}.
\label{A04}
\end{equation}
We have omitted a background integral in \eqref{A02} which is
non-leading in the high energy limit ($-z_t(s,t) \gg 1$). Note
that second term in the RHS of Eq.~\eqref{A02} is usually
discarded, since it is also negligible at $-z_t(s,t) \gg 1$.
Nevertheless, it becomes important if we look for possible
non-physical singularities.

For even signature ($\xi = +1$), one gets the real part of the
amplitude in the form:%
\footnote{Note that $\text{Im}P_l(z) = \text{Im}Q_l(z) = 0$ at $z
> 1$, for any real $l$~\cite{Erdelyi:I}.}
\begin{equation}
\text{Re} A(s,t)\Big|_{pole}  = -16 \pi^2 (2\alpha + 1) \, \beta
\left[ \frac{1 + \cos \pi \alpha}{\sin \pi \alpha} \,
P_{\alpha}(-z)- \frac{2}{\pi} \, Q_{\alpha}(-z) \right],
\label{A06}
\end{equation}
where simplified notations $\alpha \equiv \alpha(t)$, $\beta
\equiv \beta(t)$, and $z \equiv z_t(s,t)$ are introduced. In order
to analyze singularities of the amplitude in $\alpha$, it is
convenient to represent the expression in the RHS of
Eq.~\eqref{A06} via hypergeometric functions~\cite{Erdelyi:I}:
\begin{eqnarray}
&& \frac{1 + \cos \pi \alpha}{\sin \pi \alpha} \, P_{\alpha}(-z)-
\frac{2}{\pi} \, Q_{\alpha}(-z) = -\frac{\pi^{-1/2}}{\cos \pi
\alpha} \Bigg[ (-2z)^{\alpha} \, \frac{(1 + \cos \pi \alpha) \,
\Gamma(-\alpha)}{\Gamma(-\alpha +
1/2)} \nonumber \\
&& \times \, {}_2F_1 \left( - \frac{\alpha}{2}, \, -
\frac{\alpha}{2} + \frac{1}{2}; \, - \frac{\alpha}{2} + 1; \,
\frac{1}{z^2} \right) - (-2z)^{-\alpha - 1} \, \frac{(1 - \cos \pi
\alpha) \, \Gamma(\alpha + 1)}{\Gamma(\alpha +
3/2)}  \nonumber \\
&& \times \, {}_2F_1 \left( \frac{\alpha}{2} + \frac{1}{2}, \,
\frac{\alpha}{2} + 1; \, \frac{\alpha}{2} + \frac{3}{2}; \,
\frac{1}{z^2} \right) \Bigg].
\label{A08}
\end{eqnarray}
This expression is symmetric under replacement $\alpha \rightarrow
-\alpha - 1$. Note that the ratio ${}_2F_1(a,\,b;\,c;z)/\Gamma(c)$
has neither singularities nor zeros in $c$.

It follows from \eqref{A08} that the amplitude has two sets of
simple poles: physical singularities at $\alpha(t) = 2m$, $m=0, 1
\ldots$, and tachyon poles at $\alpha(t) = -(2n+1)$, $n=0, 1
\ldots$. If the term $ (2/\pi)\,Q_{\alpha}(-z)$ is disregarded in
\eqref{A06}, the tachyon poles are shifted to the points
$\alpha(t) = -2n$. Notice, there are no poles in the RHS of
\eqref{A08} at half-integer $\alpha$, since an expression in
square brackets tends to zero and cancels zeros of a function
$\cos \pi \alpha$ at these points.

Let us demonstrate that the tachyon poles are fictitious ones. In
order to do this, we will consider the
Mandelstam-Sommerfeld-Watson transformation which is based on
using of the Legendre function of the second kind~\cite{Collins}.
By disregarding all the terms but the pole contribution and the
sum in positive angular momenta, we get:
\begin{eqnarray}
A'(s,t) \! &=& 16\pi \, [1 + \xi \exp(i \pi \alpha(t))] \, \Big[
(2\alpha(t) + 1) \beta(t)
\frac{Q_{-\alpha(t) - 1}(-z_t(s,t))}{\cos \pi \alpha(t)} \nonumber \\
&-& \frac{1}{\pi} \, \sum_{l=1}^{\infty} (-1)^{l-1} (2l) \,
A_{l-1/2}^{\xi}(t) \, Q_{l-1/2}(-z_t(s,t)) \Big].
\label{A10}
\end{eqnarray}

In particular, the contribution of the Regge trajectory (with $\xi
= +1$) into the real part of the amplitude,
\begin{equation}
\text{Re} A'(s,t)\Big|_{pole}  = 16 \pi (2\alpha + 1) \, \beta \,
\frac{1 + \cos \pi \alpha}{\cos \pi \alpha} \, Q_{-\alpha -
1}(-z),
\label{A12}
\end{equation}
reveals \emph{the same} physical singularities (at $\alpha(t) =
2m$, $m=0, 1 \ldots$), as it can be easily seen from the
relation~\cite{Erdelyi:I}
\begin{eqnarray}
Q_{-\alpha - 1}(-z) &=& \pi^{1/2} \, (-2z)^{\alpha} \,
\frac{\Gamma(-\alpha)}{\Gamma(-\alpha + 1/2)} \nonumber \\
&\times& \, {}_2F_1 \left( - \frac{\alpha}{2}, \, -
\frac{\alpha}{2} + \frac{1}{2}; \, - \frac{\alpha}{2} + 1; \,
\frac{1}{z^2} \right).
\label{A14}
\end{eqnarray}

The zeros of the function $\cos \pi \alpha$ in Eq.~\eqref{A10} can
result in singularities of $\text{Re} A'(s,t)$ at half-integer
$\alpha$. However, the poles of $(\cos \pi \alpha)^{-1}$ at
$\alpha(t) = n+1/2$, with $n=0, 1 \ldots$, and corresponding poles
of the partial amplitudes in the sum in Eq.~\eqref{A10} cancel
out. To see this, one should use the formula
\begin{equation}
Q_{l-1/2}(z) = Q_{-l-1/2}(z),
\label{A16}
\end{equation}
valid for any integer $l$~\cite{Erdelyi:I}. Only tachyon poles,
$\alpha(t) = -(n+1/2)$, $n=0, 1 \ldots$, survive. Thus, we see
that positions of the tachyon poles are different if we use
Mandelstam-Sommerfeld-Watson transformation instead of standard
Sommerfeld-Watson transformation. Moreover, the singularities at
$\alpha(t) = -(n+1/2)$ do not appear in the amplitude as well, if
so-called Mandelstam symmetry of the partial
amplitudes with respect to the point $l=-1/2$ is assumed:%
\footnote{From the Gribov-Froissart representation, the Mandelstam
symmetry follows for all $A_l^{\xi}(t)$ with $l \geqslant N$,
where $N$ is a number of subtractions in a dispersion relation for
the amplitude, due to the symmetry property of
$Q_l(z)$~\eqref{A16}. The symmetry takes place in a potential
scattering (see \cite{Collins} for more details).}
\begin{equation}
A_{l-1/2}^{\xi}(t) = A_{-l-1/2}^{\xi}(t).
\label{A18}
\end{equation}
All said above indicates fictitious character of the singularities
at negative values of $\alpha(t)$.

\setcounter{equation}{0}
\renewcommand{\theequation}{B.\arabic{equation}}

\section*{Appendix B}
\label{sec:AppB}

In this Appendix we will calculate the asymptotics of the RHS of
Eq.~\eqref{220} at large value of variable $c = a b/R_g(s)$, where
$b$ is the impact parameter, and $a$ is defined in the text
\eqref{210}. The first quantity under consideration, $I_<(c)$, is
represented by the integral
\begin{equation}
I_<(c) = \int\limits_0^1 dz \, z \, J_0(cz)\,e^{-a^2 z^2}.
\label{B02}
\end{equation}
By using well-known relation between Bessel
functions~\cite{Erdelyi:II},
\begin{equation}
z^{\nu}J_{\nu - 1}(cz) = \frac{1}{c} \,\frac{d}{dz} \big[ z^{\nu}
J_{\nu}(cz)\big],
\label{B04}
\end{equation}
one can easily obtains from \eqref{B02}:
\begin{eqnarray}
I_<(c) &=& \frac{e^{-a^2}}{c} \, \sum_{m=0}^N \left( \frac{2
a^2}{c} \right)^m  J_{m+1}(c) + \left( \frac{2 a^2}{c}
\right)^{N+1} \int\limits_0^1 dz \, z^{N+2} \,
J_{N+1}(cz)\,e^{-a^2 z^2}
\nonumber \\
&=& \frac{e^{-a^2}}{c} \, \sum_{m=0}^N \left( \frac{2 a^2}{c}
\right)^m J_{m+1}(c) + \mathrm{o}(c^{-N-2})
\label{B06}
\end{eqnarray}
for any integer $N \geqslant 0$. Thus, we obtain the leading
asymptotic terms:
\begin{equation}
\left. I_<(c)\right|_{c \gg 1} = \frac{ e^{-a^2} }{c} \, \left[
J_1(c) + \frac{2 a^2}{c} \, J_2(c) \right] + \mathrm{o}(c^{-3}) .
\label{B08}
\end{equation}
Note, the sum in \eqref{B06} converges at $N \rightarrow \infty$
for any fixed $c$.

The quantity $I_>(c)$ is represented by the integral
\begin{equation}
I_>(c) = \int\limits_1^{\infty} \frac{dz}{z} \, J_0(cz) \, e^{-a^2
z^2}.
\label{B10}
\end{equation}
In order to estimate $I_>(c)$  at large $c$, it is convenient to
recast it in the form:
\begin{eqnarray}
I_>(c) &=& \lim_{\alpha \rightarrow 0} \Big[
\int\limits_0^{\infty} dz z^{\alpha -1} J_0(cz) \, e^{-a^2 z^2} -
\int\limits_0^1 dz z^{\alpha -1} J_0(cz) \, e^{-a^2 z^2} \Big]
\nonumber \\
&=& \lim_{\alpha \rightarrow 0} \Big[ \int\limits_0^{\infty} dz
z^{\alpha -1} J_0(cz) \, e^{-a^2 z^2} - \int\limits_0^1 dz
z^{\alpha -1} J_0(cz) \Big]
\nonumber \\
&-& \sum_{k=1}^{\infty} (-1)^k \frac{a^{2k}}{k!} \,
\int\limits_0^1 dz z^{2k -1} J_0(cz).
\label{B12}
\end{eqnarray}

By using table integrals from~\cite{Prudnikov:II}, we get (up to
power corrections in $\alpha$):
\begin{equation}
\left. \int\limits_0^{\infty} dz z^{\alpha -1} J_0(cz) \, e^{-a^2
z^2} \right|_{\alpha \rightarrow 0} \simeq \frac{a^{-\alpha}}{2}
\, \Big[ \Gamma \big( \frac{\alpha}{2} \big) - \gamma - \ln \Big(
\frac{c^2}{4a^2} \Big) + \mathrm{Ei} \Big(-\frac{c^2}{4a^2} \Big)
\Big],
\label{B14}
\end{equation}
where $\gamma$ is the Euler constant, and $\mathrm{Ei}(-z)$ is the
exponential integral. Analogously, one gets~\cite{Prudnikov:II}:
\begin{equation}
\left. \int\limits_0^1 dz z^{\alpha -1} J_0(cz) \right|_{\alpha
\rightarrow 0} \simeq \frac{1}{2} \, \Big[ \Gamma \big(
\frac{\alpha}{2} \big) - \gamma - \ln \Big( \frac{c^2}{4} \Big)
\Big] - \int\limits_c^{\infty} \frac{dz}{z} \,  J_0(z).
\label{B16}
\end{equation}
Eqs. \eqref{B12}-\eqref{B16} result in the expression
\begin{equation}
I_>(c) = \frac{1}{2} \, \mathrm{Ei} \Big(-\frac{c^2}{4a^2} \Big) +
\int\limits_c^{\infty} \frac{dz}{z} J_0(z) - \sum_{k=1}^{\infty}
(-1)^k \frac{a^{2k}}{k!} \, \int\limits_0^1 dz z^{2k -1} \,
J_0(cz).
\label{B18}
\end{equation}

By taking into account Eq.~\eqref{B04}, we get (for $N \geqslant
0$):
\begin{eqnarray}
\int\limits_c^{\infty} \frac{dz}{z} \, J_0(z) &=&  - \frac{1}{c}
\, \sum_{m=0}^N \, m! \, \left( \frac{2}{c} \right)^m J_{m+1}(c)
\nonumber \\
&+& 2^{N+1} \, (N+1)! \, \int\limits_c^{\infty} \frac{dz}{z^{N+2}}
\, J_{N+1}(z).
\label{B20}
\end{eqnarray}

A series, which arises in \eqref{B20} in the limit $N \rightarrow
\infty$, does not converge. To see this, one should use an
asymptotics of the Gamma-function~\cite{Erdelyi:I},
\begin{equation}
\Gamma(m) \Big|_{m \rightarrow \infty}  \simeq \exp \Big[ \Big( m
- \frac{1}{2} \Big) \ln m -m + \frac{1}{2} \ln (2\pi) \Big],
\label{B22}
\end{equation}
and an asymptotics of the Bessel function at large value of its
index (at fixed $c$)~\cite{Watson},
\begin{equation}
J_m(c) \Big|_{m \rightarrow \infty}  \simeq \exp \Big\{ m \Big[ 1
+ \ln \Big( \frac{c}{2} \Big) \Big] - \Big( m + \frac{1}{2} \Big)
\ln m  -  \frac{1}{2} \ln (2\pi) \Big\}.
\label{B24}
\end{equation}
Then one concludes that the $m$-th term in the series under
consideration tends to $(2m)^{-1}$ at $m \rightarrow \infty$. So,
the sum  in \eqref{B20} is an asymptotic one at large $N$.

As a result, we have (for $N \geqslant 0$):
\begin{eqnarray}
&& \sum_{k=1}^{\infty} (-1)^k \frac{a^{2k}}{k!} \, \int\limits_0^1
dz z^{2k -1} \, J_0(cz) = \frac{ e^{-a^2} - 1}{c} \, J_1(c)
\nonumber \\
&+& \sum_{k=1}^{\infty} (-1)^k \frac{a^{2k}}{k} \, \Big[
\frac{1}{c} \sum_{m=1}^{N} (-1)^m \, \frac{1}{\Gamma(k-m)} \left(
\frac{2}{c} \right)^m \! J_{m+1}(c)
\nonumber \\
&-& (-1)^N \left( \frac{2}{c} \right)^{N+1} \!\!
\frac{1}{\Gamma(k-N - 1)}\! \int\limits_0^1 dz z^{2k - N -2} \,
J_{N+1}(cz) \Big].
\label{B26}
\end{eqnarray}
Taking into account that $1/\Gamma(-n)=0$ for any non-negative
integer $n$, we come to the formula
\begin{eqnarray}
&& \sum_{k=1}^{\infty} (-1)^k \frac{a^{2k}}{k!} \, \int\limits_0^1
dz z^{2k -1} \, J_0(cz) = \frac{ e^{-a^2} - 1}{c} \, J_1(c)
\nonumber \\
&-& \frac{1}{c} \sum_{m=1}^{N} \, \left( \frac{2}{c} \right)^m \!
\gamma(m+1,a^2) \, J_{m+1}(c)
\nonumber \\
&-& \left( \frac{2}{c} \right)^{N+1} \, \int\limits_0^1 dz z^{- N
-2} \, \gamma(N+2, z^2 a^2) \, J_{N+1}(cz) \Big].
\label{B28}
\end{eqnarray}
Here $\gamma(a,x)$ is the incomplete Gamma-function. At $N
\rightarrow \infty$, the sum in the RHS of Eq.~\eqref{B28}
converges. It can be easily shown if we use the expansion of the
incomplete Gamma-function, $\gamma(m+1,1) = \sum_{k=0}^m \,
[k!(k+m+1)]^{-1}$, and use \eqref{B24}.

It follows from Eqs.~\eqref{B18}, \eqref{B20} and \eqref{B28} that%
\footnote{Notice, the exponetial integral $\mathrm{Ei}(-c^2/4a^2)
$ decreases exponentially at $c \rightarrow \infty$.}
\begin{eqnarray}
I_>(c) &=& - \frac{ e^{-a^2} }{c} \, J_1(c) - \frac{1}{c} \,
\sum_{m=1}^N \, \left( \frac{2}{c} \right)^m m! \, J_{m+1}(c)
\nonumber \\
&+& \frac{1}{c} \, \sum_{m=1}^{\infty} \, \left( \frac{2}{c}
\right)^m \gamma(m+1,a^2) \, J_{m+1}(c) + \mathrm{o}(c^{-N-2}),
\label{B30}
\end{eqnarray}
and we obtain the leading asymptotic terms:
\begin{equation}
I_>(c) \Big|_{c \gg 1} = - \frac{ e^{-a^2} }{c} \, \left[ J_1(c) +
\frac{2(1+a^2)}{c} \, J_2(c) \right] + \mathrm{o}(c^{-3}).
\label{B32}
\end{equation}
Notice that the leading terms in \eqref{B08} and \eqref{B32}
(proportional to $J_1(c)$) cancel out. Thus, we get
\begin{equation}
[I_<(c) + I_>(c)] \Big|_{c \gg 1} \simeq - \frac{ 2e^{-a^2} }{c^2}
\, J_2(c),
\label{B34}
\end{equation}
and we arrive at the asymptotics presented in the text
\eqref{260}. The complete asymptotic expansion of the functions
$I_<(c)$ and $I_>(c)$ can be obtained if desired from
Eqs.~\eqref{B06} and \eqref{B30}, respectively.

\setcounter{equation}{0}
\renewcommand{\theequation}{C.\arabic{equation}}

\section*{Appendix C}
\label{sec:AppC}

In order to find $\mathrm{Re} \tilde{A}(s,t)$, we need to
calculate the integral
\begin{equation}
M = \int\limits_0^{\infty} db \, b \, J_0(q_{\bot} b) \Big[ I_<
\left( \frac{a b}{R_g(s)} \right) + I_> \left( \frac{a b}{ R_g(s)}
\right) \Big].
\label{C02}
\end{equation}
By taking into account formulae from the Appendix~B, one can get:
\begin{eqnarray}
I_<(c) + I_>(c) &=& - \frac{ 2e^{-a^2} }{c^2} \, J_2(c) + \left(
\frac{2 a^2}{c} \right)^{2} \int\limits_0^1 dz \, z^{3} \,
J_{2}(cz)\,e^{-a^2 z^2}
\nonumber \\
&+& \left( \frac{2}{c} \right)^{2} \, \int\limits_0^1
\frac{dz}{z^3} \, \gamma(3, z^2 a^2) \, J_{2}(cz)
\nonumber \\
&+& 8 \, \int\limits_c^{\infty} \frac{dz}{z^{3}} \, J_{2}(z) +
\frac{1}{2} \, \mathrm{Ei} \Big(-\frac{c^2}{4a^2} \Big),
\label{C04}
\end{eqnarray}
where again we have defined $c=a b/ R_g(s)$.

By the use of Eq.~\eqref{B04} and other relation between the
Bessel functions~\cite{Erdelyi:II},
\begin{equation}
\frac{d}{dz} \big[ z^{-\nu} J_{\nu}(cz)\big] = - c z^{-\nu}
J_{\nu+1}(z),
\label{C08}
\end{equation}
the following formula can be derived ($\alpha, \, \beta > 0$):%
\footnote{A special case of this equation for $\mu = 0$ and $a=b$
can be found in \cite{Prudnikov:II}.}
\begin{equation}
\int\limits^x \frac{dz}{z^{\mu}}  J_{\nu}(\alpha z)
J_{\nu+\mu+1}(\beta z) = \frac{1}{\beta x^{\mu}} \,
\sum_{m=1}^{\infty} \Big( \frac{\beta}{\alpha} \Big)^m J_{\nu +
m}(\alpha x)
 J_{\nu+\mu+m}(\beta x) + \mathrm{const}.
 \label{C10}
\end{equation}
In particular, one obtains (for $A, \, q_{\bot} > 0$):
\begin{eqnarray}
&& \int\limits_{0}^{\infty} \frac{db}{b} \, J_0(q_{\bot} b) \, J_2
(A b)
\nonumber \\
&& = \frac{1}{A \, b} \, \sum_{m=1}^{\infty} \Big(
\frac{A}{q_{\bot}} \Big)^{m} J_m \big( q_{\bot} b \big) \,
J_{m+1}( A b) \, \Big|_{b=0}^{b=\infty} = 0.
\label{C12}
\end{eqnarray}
Thus, first three terms in the RHS of Eq.~\eqref{C04} gives zero
after the integration in variable $b$ in \eqref{C02}.

The next to the last term in Eq.~\eqref{C04} results in
\begin{eqnarray}
&& 8 \int\limits_0^{\infty} db \, b \, J_0(q_{\bot} b)
\int\limits_c^{\infty} \frac{dz}{z^{3}} \, J_{2}(z)
\nonumber \\
&& = \frac{8 R_g(s)}{a q_{\bot}} \int\limits_0^{\infty}
\frac{dz}{z^{2}} \, J_{2}(z) \, J_1 \Big[ \frac{R_g(s)q_{\bot}}{a}
z \Big].
\label{C14}
\end{eqnarray}
The integral in \eqref{C14} is a table one (see
\cite{Prudnikov:II}):
\begin{eqnarray}
&& \int\limits_0^{\infty} \frac{dz}{z^{2}} \, J_{2}(z) \, J_1
\Big[ \frac{R_g(s)q_{\bot}}{a} z \Big]
\nonumber \\
&& = \left\{
\begin{array}{cc}
  \displaystyle
  \frac{R_g(s) q_{\bot}}{8a} \, \Big[ 2 - \frac{R_g^2(s) \,
  q_{\bot}^2}{a^2} \Big], & \mathrm{for} \quad |t| < a^2
  R_g^{-2}(s) \\ \\
  \displaystyle
  \frac{a}{8 R_g(s) \, q_{\bot}} \, , & \mathrm{for} \quad
  |t| \geqslant a^2 R_g^{-2}(s). \\
\end{array}
\right.
\label{C16}
\end{eqnarray}

The contribution from the last term in Eq.~\eqref{C04} can be also
explicitly calculated with the help of a table integral from
\cite{Prudnikov:II}:
\begin{equation}
\frac{1}{2} \, \int\limits_0^{\infty} db \, b \, J_0(q_{\bot} b)
\, \mathrm{Ei} \Big(-\frac{c^2}{4a^2} \Big) = -
\frac{1}{q_{\bot}^2} \, \Big[ 1 - e^{- q_{\bot}^2 R_g^2(s)} \Big],
\label{C18}
\end{equation}
and we get:
\begin{eqnarray}
&& M \, \Big|_{q_{\bot} < \, a R_g^{-1}(s)} = \frac{R_g^2(s)}{a^2}
\, \Big[ 2 - \frac{R_g^2(s) \, q_{\bot}^2}{a^2} \Big] -
\frac{1}{q_{\bot}^2} \, \Big[ 1 - e^{- q_{\bot}^2 R_g^2(s)} \Big],
\nonumber \\
&& M \, \Big|_{q_{\bot} \geqslant \, a  R_g^{-1}(s)} =
\frac{1}{q_{\bot}^2} \; e^{- q_{\bot}^2 R_g^2(s)}.
\label{C20}
\end{eqnarray}
As a result, we arrive at the expression for the amplitude
presented in the text~\eqref{400}.

\end{document}